\documentstyle[epsfig, twocolumn]{venice97}

\begin{document}

%% Do not remove the following six lines:
\setlength{\parindent}{0pt}
\setlength{\parskip}{ 10pt plus 1pt minus 1pt}
\setlength{\hoffset}{-1.5truecm}
\setlength{\textwidth}{ 17.1truecm }
\setlength{\columnsep}{1truecm }
\setlength{\columnseprule}{0pt}
\setlength{\headheight}{12pt}
\setlength{\headsep}{20pt}
\pagestyle{veniceheadings}

%% Title - should be in capitals:
\title{\bf THE LUMINOSITIES AND DIAMETERS OF MIRA VARIABLES FROM HIPPARCOS
PARALLAXES \thanks{Based on data from the Hipparcos astrometry satellite}}

%% If the author list spans more than one line then the {\bf (bold
%% font)} command must be inserted for each line
\author{{\bf  P.A.~Whitelock$^1$, F.~van Leeuwen$^2$, M.W.~Feast$^3$}
\vspace{2mm} \\
$^1$South African Astronomical Observatory, PO Box 9, Observatory, 7935,
South Africa\\
$^2$Royal Greenwich Observatory, Madingley Rd, Cambridge, CB3 0EZ,
England\\
$^3$Astronomy Department, University of Cape Town, Rondebosch, 7700,
South Africa}

\maketitle

\begin{abstract}
   Hipparcos trigonometrical parallaxes of Mira-type variables have
been combined with ground-based angular diameter measurements to derive 
linear diameters. Of eight stars with ground-based data, six have diameters 
indicating overtone pulsation whilst two, both with periods over 400 day, are
pulsating in the fundamental.\\

  Hipparcos parallaxes of 11 Miras have been combined with extensive infrared
photometry to determine the zero-point of the Mira period-luminosity
relation. Adopting the relation at K (2.2$\mu$m), since this is less likely
to be subject to abundance effects than that at $M_{bol}$, leads to a distance
modulus for the LMC of 18.6 mag with a uncertainty of slightly less than
0.2 mag.\\

  A brief discussion is given of the preliminary analysis of the parallaxes
of a much larger sample of Miras. Some consideration is given to possible
problems in interpreting the Hipparcos data which arise because of the
physical characteristics of the Mira variables.  Finally the apparent
low-luminosity of the carbon Mira, R Lep, implied by the Hipparcos results
leads to an interesting problem in AGB evolution.
  \vspace {5pt} \\

%% Do not remove the previous commands. Your abstract should 
%% end with \vspace {5pt} \\  

%% Please insert your keywords here.
  Key~words: space astrometry; Mira variables; carbon stars; distance scale;
angular diameters; infrared photometry; asymptotic giant branch.

\end{abstract}

\section{INTRODUCTION}

Miras are cool, long period, large amplitude variables.  Their spectra are
dominated by molecular absorption features and show hydrogen emission lines,
as a result of shock waves, during certain phases of their pulsation cycles.
Their periods are longer than 100 days and their visual light amplitudes in
excess of 2.5 mag. The stars whose parallaxes are discussed below have
visual light variations, according to the General Catalogue of Variable
Stars, of 5 to 10 mag, although their bolometric amplitudes are less than
one mag. Examples of this type of star are found in the metal-rich globular
clusters where they are the most luminous stars present. The Mira
evolutionary phase is short lived, about $2\times 10^5$ year, and is
accompanied by heavy mass loss, typically between $10^{-6}$ and 
$10^{-5}$ $M_{\odot}\,{\rm yr^{-1}}$. We therefore understand the Miras to
be at the very tip of the Asymptotic Giant Branch, probably in the so-called
``superwind phase''. The Miras have potential as distance indicators among
old metal-rich populations as they represent the maximum luminosity phase of
such stars. They are best studied in the near-infrared as their flux
distribution peaks between 1 and 2 $\mu$m.

\section{THE HIPPARCOS DATA}

The authors have Hipparcos data on 180 Mira-like variables, but most of
these
have only been analysed in a superficial manner. The bulk of this paper,
therefore, concerns 16 Miras for which early access to Hipparcos data was
obtained and which have been examined in detail. These stars were selected,
prior to any knowledge of their parallaxes, as potentially interesting
objects. The details of this analysis are given by van Leeuwen et al.\
(1997).  The sample of 16 comprises: 10 normal Miras of spectral type M, $o$
Cet, R Tri, R Hor, U Ori, R Car, R Leo, S Car, RR Sco, T Cep and R Cas; one
S-type, $\chi$ Cyg; one C-type, R Lep; two M-types with slowly decreasing
periods, R Hya and R Aql, which may be undergoing He-shell flashes (Wood \&
Zarro 1981); one M-type star which has a double period, R Cen; the
oxygen-rich symbiotic Mira, R Aqr, which is weakly interacting with a white
dwarf and has been well studied at many wavelengths.\\

The Hipparcos parallaxes ($\pi$) and their errors ($\sigma_{\pi}$) are given
in Table 1, together with the interstellar extinction at V ($A_V$), the
extinction corrected mean 2.2$\,\mu$m magnitude ($\bar{K}_0$), the number of
$JHK$ measurements (N) and the mean apparent bolometric magnitude
($\bar{m}_{bol}$).  For the analysis a weighted mean of the Hipparcos
parallax and that measured by Gatewood (1992) was used for R Leo. The mean
bolometric magnitudes were derived from near-infrared ($JHK$) photometry.
For most of the Miras extensive observations have been made from the South
African Astronomical Observatory (SAAO), but R
Cas and T Cep are inaccessible from the south so for these stars data
obtained in the Crimea was provided by Boris Yudin.

\begin{table*}
  \begin{center}
    \leavevmode
    \footnotesize
\caption{\em Observational Data}\vspace*{1mm}
\begin{tabular}[h]{@{}lcccrrcccc}
      \hline \\[-5pt]
\multicolumn{1}{c}{Star} & \multicolumn{1}{c}{$\pi$} & 
\multicolumn{1}{c}{$\sigma_{\pi}$} & \multicolumn{1}{c}{$ A_{V}$} & 
\multicolumn{1}{c}{$ \bar{K}_0$} & \multicolumn{1}{c}{N} & 
\multicolumn{1}{c}{$ \bar{m}_{bol}$} &
\multicolumn{1}{c}{$\varphi$} & \multicolumn{1}{c}{$\sigma_{\varphi}$} & 
\multicolumn{1}{c}{$\log$P}\\ 
 & \multicolumn{2}{c}{(mas)} & \multicolumn{2}{c}{(mag)} &
& (mag) & \multicolumn{2}{c}{(mas)} & (P in day)\\
      \hline \\[-5pt]
o Cet & 7.79 & 1.07 & 0.01 & --2.50 & 98 & 0.70 & 33.6 & 3.5 & 2.521\\
R Tri & 2.51 & 1.69 & 0.14 & 0.93 & 9 & 4.04 & & & 2.426\\
R Hor & 3.25 & 1.08 & 0.02 & --0.94 & 41 &2.22 & & & 2.611\\
R Lep & 3.99 & 0.85 & 0.08 & --0.01 & 121 & 3.45 & & & 2.630\\
U Ori & 1.52 & 1.65 & 0.23 & --0.64 & 50 & 2.54 & 18.5 & 2.6 & 2.566\\
R Car & 7.84 & 0.83 & 0.13 & --1.35 & 77 & 1.74 & & &  2.490\\
R Leo & 9.87 & 2.07 & 0.02 & --2.55 & 50 & 0.69 & 37.4 & 2.3 & 2.491\\
S Car & 2.47 & 0.63 & 0.35 & 1.84 & 34 & 4.65 & & & 2.175\\
R Hya & 1.62 & 2.43 & 0.03 & --2.48 & 51 & 0.66 & 28.7 & 3.3 & 2.590\\
R Cen & 1.56 & 0.84 & 0.21 & --0.72 & 67 & 2.38 & & & 2.737\\
RR Sco & 2.84 & 1.30 & 0.20 & --0.25 & 55 & 2.88 & & & 2.449\\
R Aql & 4.73 & 1.19 & 0.23 & --0.78 & 45 & 2.34 & 17.5 & 3.7 & 2.454\\
$\chi$ Cyg & 9.43 & 1.36 & 0.14 & --1.93 & 11 & 1.39 & 28.9 & 3.0 & 2.611\\
T Cep & 4.76 & 0.75 & 0.11 & --1.71 & 2 & 1.50 & 24.3 & 4.4 & 2.589\\
R Aqr & 5.07 & 3.15 & 0.01 & --1.02 & 120 & 2.26 & & & 2.588\\
R Cas & 9.37 & 1.10 & 0.12 & --1.80 & 13 & 1.40 & 24.9 & 2.9 & 2.633\\
      \hline \\[-5pt]
\end{tabular}
\end{center}
\end{table*}

\begin{figure*}
  \begin{center}
    \leavevmode
  \epsfbox[76 101 390 308]{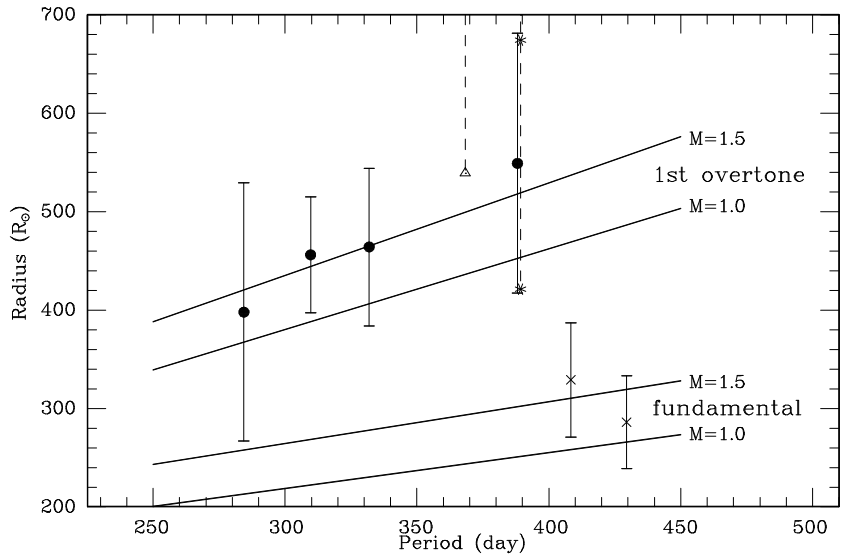}
  \end{center}
  \caption{\em The Period-Radius relation for Mira variables. Filled
circles: R Aql, R Leo, $o$ Cet, T Cep; crosses: $\chi$ Cyg, R Cas. 1-$\sigma$
error bars are shown. The triangle is the 1-$\sigma$ lower limit for U Ori
and the asterisks are the 1-$\sigma$ and 2-$\sigma$ lower limits for R Hya.
The lines are the predicted relations for fundamental and for first overtone
pulsation.} \end{figure*}

\section{STELLAR RADII}

For eight of the Miras with parallaxes Haniff et al.\ (1995) have measured
angular diameters using optical interferometric techniques.  These ($\phi$)
are listed in Table 1, together with their errors ($\sigma_{\phi}$). Thus
their linear radii can be calculated and compared with theoretical values.
Such a comparison is shown in Figure 1 as a function of pulsation period.
The theoretical predictions are shown for stars of mass 1.0 and
1.5~$M_{\odot}$ pulsating in the fundamental and first overtone modes. The
error bars on the observations represent the combined error on the radii and
on the parallaxes.\\

The pulsation mode of the Miras has been a source of controversy for many
years with good reasons for suggesting that the observed mode is the first
overtone and other, also good, arguments for suggesting it must be the
fundamental. Feast (1996) and Bessell et al.\ (1996) recently summarised the
situation from very different viewpoints. Figure 1 can be interpreted
as showing that most Miras with periods less than 400 day are pulsating
in an overtone, while at least some of the stars with longer periods
are pulsating in the fundamental.

\section{PERIOD-LUMINOSITY RELATION}

Period-luminosity (PL) relations have been established for Miras in the 
Large Magellanic Cloud (LMC)
(Feast et al.\ 1989).  These are of the form:
\begin{eqnarray}
M_K & = & -3.47 \log P + \beta_1,\\
M_{bol} & = & -3.00 \log P + \beta_2,
\end{eqnarray}
 where $\beta_1$ and $\beta_2$ are the zero-points which depend on the
distance modulus of the LMC.  These relations show only a small scatter of
0.13 mag in $K$ and 0.16 mag in $m_{bol}$. It is with them that we compare
the Hipparcos results. The relatively high errors on the parallaxes of the
local Miras prevent an unbiased full solution for the period-luminosity
relation. The slope of the PL relation is therefore assumed to be the same
as in LMC, as given in equations 1 or 2, and the zero-point is derived from
the new data. The equations are solved in the following form, weighted
according to the inverse square of the error on the parallax:
 \begin{eqnarray}
10^{0.2 \beta_1} & = & 0.010 \,\pi\, 10^{0.2(3.47 \log P +K_0)},\\
10^{0.2 \beta_2} & = & 0.010 \,\pi\, 10^{0.2(3.00 \log P +m_{bol})}.
\end{eqnarray}
 11 stars are used in the solution while the following are omitted: the two
fundamental pulsators, $\chi$ Cyg and R Cas, the carbon star, R Lep, the double
period star, R Cen, and the symbiotic, R Aqr. The following zero-points, and 
corresponding distance moduli for the LMC, were derived:\\

$K:\ \ \ \ \beta_1=0.88\pm 0.18\  {\rm mag} \\ 
\hspace*{1cm} LMC\ (m-M)_0=18.60\ {\rm mag} $\\

$M_{bol}:\  \beta_2=2.88\pm 0.17\  {\rm mag} \\
\hspace*{1.1cm}  LMC\ (m-M)_0=18.47\ {\rm mag}$\\

 The K solution is probably to be prefered as it appears to be less subject
to differences in metallicity (e.g.\ Feast 1996). The derived LMC distance
modulus of $18.60\pm0.18$ is in good agreement with Feast \& Catchpole's
(1997) value of $18.70\pm0.10$ derived from the Cepheid variables, using a
new calibration based on Hipparcos results.\\

The $M_{bol}$ PL diagram is illustrated in Figure 2 where it can be seen
that the two fundamental pulsators and the carbon star, R Lep, lie distinctly
below the PL relation derived above. The K PL diagram, which is similar to
Figure 2, is not illustrated here, but is shown in van Leeuwen et al.\ (1997).

\begin{figure*}[!ht]
  \begin{center}
    \leavevmode
  \epsfbox[61 91 416 309]{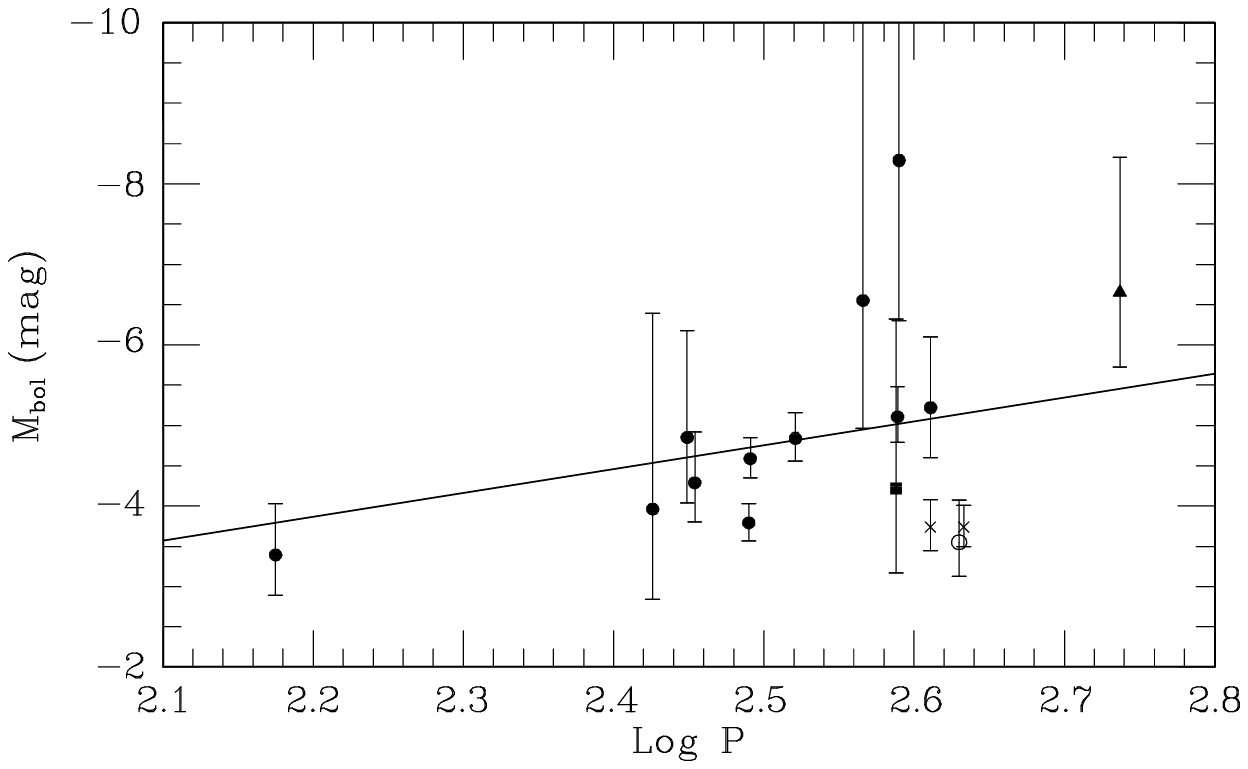}
  \end{center}
  \caption{\em The $M_{bol}$ - $\log P$ relation. Symbols: open circle: the
carbon Mira, R Lep; filled square: the symbiotic Mira, R Aqr; filled
triangle: the double period Mira, R Cen; crosses: the fundamental pulsators,
$\chi$ Cyg and R Cas; filled circles the other Miras. The line represents
the period-luminosity relation for the LMC assuming a distance modulus of
18.60 mag} \end{figure*}

\section{LARGER DATA SET}

The larger data set includes parallaxes and proper motions for 180 stars.
Of these there are 115 oxygen- and 19 carbon-rich Miras for which there is
also infrared photometry, either from SAAO or from the Crimea.  It would be
extremely useful to have infrared photometry for more of the northern stars;
such measurements require a telescope with an aperture no more than 0.5-m.
Even among the stars for which measurements have been made the light-curve
coverage is often poor with only one or two observations per star. Note that
rather few of these stars have parallaxes which are larger than their errors
and quite a number of them have negative parallaxes, so the data set is not
obviously a vast improvement over that discussed above and by van Leeuwen et
al.\ (1997).\\

There is little purpose in listing numbers from a preliminary analysis
that will undoubtedly be superseded in the near future. However, two 
points can usefully be made here. First, the PL zero-point estimated from the
103 oxygen-rich Miras with periods less than 400 days is very close to
that derived for the 11 stars discussed above. Secondly, the zero-point
derived from the 19 carbon-rich Miras is about 0.7 mag fainter than 
that derived from the short-period oxygen-rich stars. Thus R Lep is not 
the only carbon-rich Mira which lies below the PL relation. The consequences
of this are discussed further in section 7.

\section{POSSIBLE PROBLEMS WITH THE PARALLAXES}

Several Miras had a significant fraction of data rejected when the
astrometric solutions were made. This can be seen from the parameter
tabulated in column 29 of the Hipparcos catalogue (H29).  Among the stars
discussed above are R Leo, $\chi$ Cyg and R Car, which had 10 percent or
more data rejected. Several Miras have a rather poor ``goodness of fit
parameter" as specified in column 30 of the catalogue (H30), although $\chi$
Cyg is the only one among those discussed above.\\

$\chi$ Cyg is one of the fundamental pulsators with a faint bolometric
magnitude. R Car is also faint in Figure 2 compared to the other stars
although it was included in the solution for the PL zero-point. On the other
hand the Hipparcos parallax of R Leo agrees well with that already published
by Gatewood (1992). It is therefore not clear that the H29 and H30
parameters are indicating any real problem with the measured parallax.  They
may just indicate what we anticipated before the satellite flew, i.e. that
it would be difficult for Hipparcos to measure some stars near minimum light
and we should therefore expect some data rejection for these large amplitude
pulsators. $\chi$ Cyg is particularly faint, reaching minimum at $Hp=12.09$
mag as listed in the Hipparcos catalogue. In the next stage of the analysis
we propose to avoid most of the minimum light problems by making a direct
absolute magnitude calibration using the intermediate astrometric data as
described by van Leeuwen (this symposium) and by van Leeuwen \& Evans
(1997). The minimum-light observations will be outweighed in such a solution
by the other measurements. The common solution also offers better
possibilities for recognising faulty observations.\\

Another problem is possible temporal changes in the light distribution
across the disk of the Mira.  The parallaxes of these Miras are actually
smaller than their angular diameters. The parallaxes range from about 2 to
10 milli-arcsec, while the diameters of the eight which have been measured
are between 17 and 34 milli-arcsec. We know that the Miras do not present
uniform stellar disks; observations of R Leo made with the Fine Guidance
Sensor on the HST, in two orthogonal directions, show that one axis exceeds
the other by 11 percent and even bigger differences are found in some stars
(Lattanzi et al.\ 1997). The cause of this is unknown as is the way it
changes with time. It might be non-radial pulsation or it might be star
spots resulting from large convection cells. Thus it is not yet clear if the
effect will give us systematic errors in the parallax or increase the
uncertainty of the measurement or even if it will have any effect at all. It
is, however, clearly important to obtain more data on possible asymmetries
and to investigate any possible effect on the parallaxes.\\

With the caveat that all these things need looking into in more detail, we
do not have any strong reasons to feel there are any particular problems
with the Hipparcos parallax data. In fact the Hipparcos parallaxes represent
a remarkable set of data which will put our understanding of every aspect of
the distance scale onto a new and firmer footing.

\section{THE CARBON RICH MIRAS}

The Hipparcos parallax of R Lep presents a problem which is discussed here
very briefly. The parallax and infrared photometry imply a bolometric
magnitude of $-3.6^{-0.52}_{+0.42}$ mag ($L=2.2 \times 10^3\ L_{\odot}$).
This luminosity is faint even for the transition luminosity from oxygen to
carbon stars, which in the LMC ranges from around $-3.8$ to about $-4.8$,
depending on mass and metallicity.  Fainter transition luminosities are
found only in the most metal-weak populations. Although we do not know the
metallicity of R Lep, it has strong lines and seems unlikely to be metal
deficient. In any case, stars near the transition luminosity are not
normally large amplitude variables. Miras are normally a magnitude or so
brighter; nearer the $-5.1$ mag ($L=9\times10^3\ L_{\odot}$) predicted for R
Lep from the PL relation. As mentioned in section 5 the parallaxes of the
other carbon Miras also imply faint absolute magnitudes for at least some of
the stars. \\

We are therefore drawn to the possibility that R Lep actually has a
luminosity close to that predicted from the PL relation, but that we have
underestimated its apparent luminosity for some reason. For example such a
situation could arise if R Lep was undergoing asymmetric mass-loss, and we
were viewing the star through a dusty equatorial torus. This possibility has
been discussed in more detail by Whitelock et al.\ (1997) who present
infrared light-curves for R Lep and other carbon Miras. Several of these
stars, including R Lep, have erratic long-term changes in their output which
are reminiscent of the light-curves of R Corona Borealis (RCB) stars. This
has been interpreted as the result of dust formation following non-periodic
mass loss from the carbon Miras. If these stars eject dust in random
directions, as the RCB stars apparently do, it will not explain the
faintness of R Lep. If, however, they eject it preferentially in particular
directions, such as in the equatorial plane, and it happens that we are
viewing R Lep from such a preferential direction then its faintness might be
understood as a consequence of circumstellar obscuration. On the other hand
if some of the carbon Miras are actually considerably fainter than was
Previously thought then it will have repercussions for our understanding of
dredge-up and of AGB evolution. Clearly this needs further investigation.

\section*{ACKNOWLEDGEMENTS}
We are grateful to Boris Yudin of the Sternberg Astronomical Institute,
and to various members of the South African Astronomical Observatory  
for use of their infrared photometry of Mira variables prior to
publication.

\end{document}